\documentclass[aps,prd,preprint,titlepage,raggedbottom,amsmath,amssymb,superscriptaddress,floatfix]{revtex4-2}
\usepackage{graphicx}
\usepackage{bm}
\usepackage[colorlinks=true,citecolor=blue,urlcolor=blue,linkcolor=blue]{hyperref}
\graphicspath{{figures/}}

\begin{document}
\title{Quantum Autoencoder on Real Hardware for New Physics Searches at Muon Colliders}
\author{Shuai Zhang}
\author{Ji-Chong Yang}
\email{yangjichong@lnnu.edu.cn}
\affiliation{Department of Physics, Liaoning Normal University, Dalian 116029, China}
\affiliation{Center for Theoretical and Experimental High Energy Physics, Liaoning Normal University, Dalian 116029, China}
\date{November 17, 2026}
\begin{abstract}
We apply a quantum autoencoder to search for new physics~(NP) in the tri-photon process $\mu^+\mu^-\to\gamma\gamma\gamma$ at muon colliders. Using a six-qubit circuit on the Tianyan176 quantum computer, we measure the reconstruction losses for Standard Model~(SM) events and anomalous quartic gauge couplings~(aQGCs) events. The results demonstrate that the hardware-implemented quantum autoencoder effectively discriminates NP signals from the SM background. Concurrently, we derive the expected constraints on the aQGCs coefficients and compare them with those obtained from conventional methods.
\end{abstract}
\maketitle

\section{Introduction}
Searches for physics beyond the Standard Model~(SM) require methods that distinguish rare signals from abundant backgrounds. Autoencoders learn a compressed representation of reference events and assign a high anomaly score to events that reconstruct poorly. A quantum autoencoder~(QAE) implements this compression using parameterized quantum circuits~\cite{Romero_2017}. Its utility for collider analysis depends on whether this anomaly score retains sufficient discriminating information on noisy quantum hardware.

Quantum anomaly detection has been explored in high energy physics~(HEP) through both simulations and hardware implementations~\cite{Ngairangbam2022,Belis2024,Duffy2025}. For instance, an earlier hardware study by Ngairangbam et al.~\cite{Ngairangbam2022} used the fidelity of discarded qubits to distinguish heavy-Higgs signals from top-quark backgrounds. Here, we reconstruct the encoded event states on a quantum processor and test whether the measured reconstruction loss separates simulated SM events from anomalous quartic gauge couplings~(aQGCs) events in the process $\mu^+\mu^-\to\gamma\gamma\gamma$.

We encode the four-momenta of the three hardest photons into a four-qubit state and evaluate the reconstruction score on Tianyan176 using a six-qubit circuit. At each collision energy, we evaluate $500$ SM events and $500$ new physics~(NP) events with $2000$ shots per circuit. The resulting measured areas under the receiver operating characteristic~(ROC) curves are presented for the different c.m. energies.

The remainder of this paper is organized as follows.
Section~\ref{sec:2} briefly introduces the physics benchmark and event samples.
Section~\ref{sec:3} details the process of quantum reconstruction.
The hardware discrimination results are presented in Section~\ref{sec:4}.
Section~\ref{sec:5} presents and compares the expected coefficient constraints of two methods.
Finally, Section~\ref{sec:6} summarizes the main conclusions.

\section{\label{sec:2}Physics benchmark and event samples}
The tri-photon process probes neutral anomalous couplings $\gamma\gamma\gamma Z/\gamma$, induced by di\-men\-sion\nobreakdash-8 operators~\cite{Eboli:2006wa,Eboli:2016kko,Yang:2020rjt}. We describe these contributions with the effective Lagrangian
\begin{equation}
\mathcal L=\mathcal L_{\rm SM}+\sum_i c_i O_{T_i},\qquad
c_i\equiv f_{T_i}/\Lambda^4,
\label{eq:eft}
\end{equation}
where each transverse operator $O_{T_i}$ contains four electroweak field-strength tensors. The Wilson coefficients $c_i$ are quoted in units of $\mathrm{TeV}^{-4}$. Since the contributions of $O_{T_1}$ and $O_{T_6}$ are equivalent to those of $O_{T_0}$ and $O_{T_5}$ for this process~\cite{Yang:2020rjt}, our phenomenological scan focuses on $O_{T_{0,2,5,7,8,9}}$ operators. The hardware benchmark excludes SMEFT interference, which is included in the cross-section scan described below. Appendices~\ref{sec:operators} and~\ref{sec:samples} give the operator definitions and sample preparation details.

Events are generated using Monte Carlo with the help of \texttt{MadGraph5\_aMC@NLO}~\cite{Alwall:2014hca}, followed by a muon collider-like detector simulation with \texttt{Delphes}~\cite{deFavereau:2013fsa}. The generation cuts are $p_{T,\gamma}>10$~GeV, $|\eta_\gamma|<2.5$, and $\Delta R_{\gamma\gamma}>0.4$. We retain events with at least three reconstructed photons and form the 12-dimensional input vector $\mathbf{x}_n$ by concatenating the four-momenta of the three highest-energy photons in descending order.

For each c.m. energy, candidate QAEs are trained on $1024$ SM events. Model selection utilizes a separate sample of $10\,000$ SM and $10\,000$ NP events, and the selected model is then evaluated on an independent validation sample of equal size. For the hardware measurements, a single random sample of $500$ events per class is drawn from a separate test set of $10\,000$ events per class.

\section{\label{sec:3}Quantum reconstruction score}

\begin{figure}[htbp]
\centering
\includegraphics[width=0.84\linewidth]{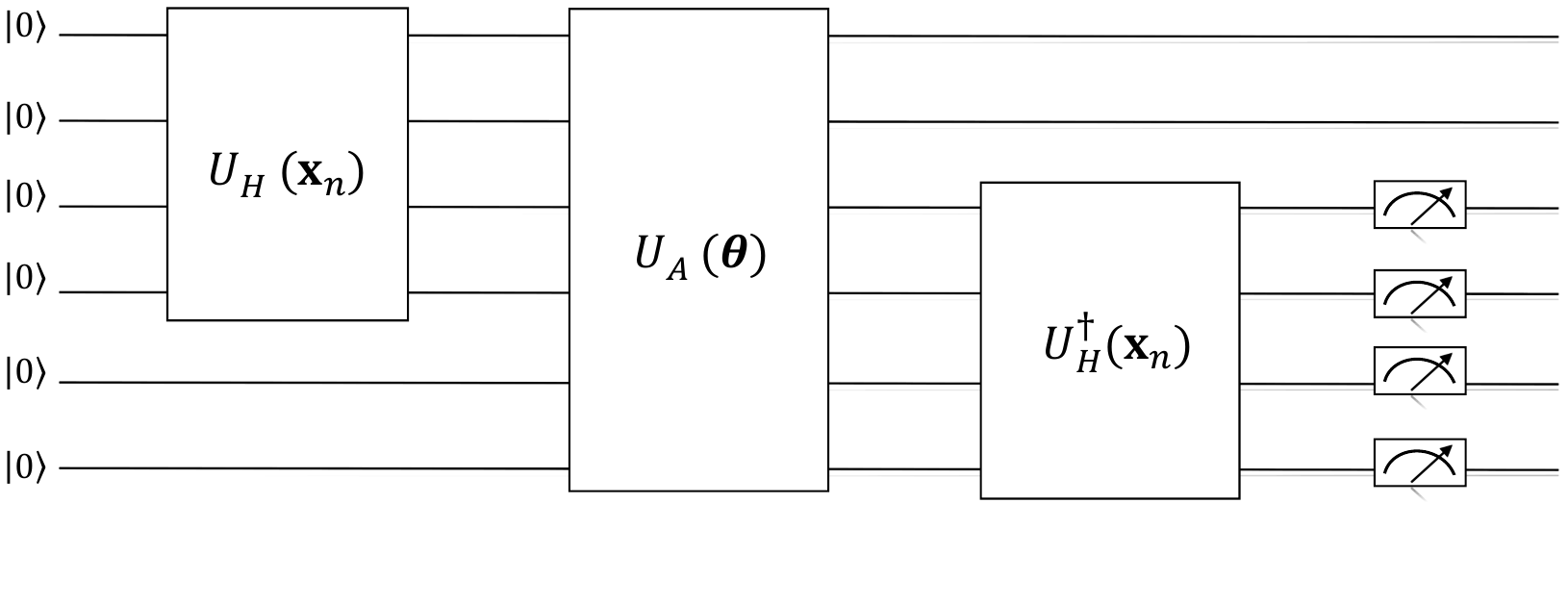}
\caption{Six-qubit circuit for measuring the reconstruction fidelity.
The data encoding $U_H(\mathbf{x}_n)$ prepares the four-qubit input
state. The QAE operation $U_A(\boldsymbol{\theta})$ acts on this
register and two additional qubits initialized in $\lvert 00\rangle$.
In the ideal circuit, after applying $U_H^\dagger(\mathbf{x}_n)$
to the four reconstructed output qubits, the probability of measuring
$0000$ on those qubits equals the reconstruction fidelity $f_n$.
The other two qubits are left unmeasured.}
\label{fig:circuit}
\end{figure}

A shallow hardware-efficient circuit prepares $|\psi_n\rangle=U_H(\mathbf{x}_n)|0000\rangle$. The QAE compresses and reconstructs this state using a parameterized circuit, as illustrated in Fig.~\ref{fig:circuit}. Because discarded degrees of freedom and device noise can leave the reconstructed register mixed, its state is denoted by $\rho_n^{\rm rec}$. The fidelity with the pure encoded input and the anomaly score are
\begin{equation}
 f_n=\langle\psi_n|\rho_n^{\rm rec}|\psi_n\rangle,\qquad l_n=1-f_n.
\label{eq:score}
\end{equation}
This expression reduces to $|\langle\psi_n|\phi_n\rangle|^2$ when the reconstructed state is pure. After applying the inverse data encoding $U_H^\dagger(\mathbf{x}_n)$, the ideal probability of obtaining $0000$ on the four reconstructed output qubits equals the reconstruction fidelity. For event $n$, we estimate the reconstruction loss as $\widehat l_n=1-n_{0000}/N_{\rm shot}$, where $n_{0000}$ is the number of all-zero outcomes and $N_{\rm shot}=2000$. We use the raw counts without error mitigation or postselection.

During training, the encoded SM events are represented by the joint state
\begin{equation}
|\Psi\rangle=\frac{1}{\sqrt{M}}\sum_{n=0}^{M-1}|\psi_n\rangle|n\rangle.
\label{eq:joint}
\end{equation}
We train a set of candidate QAEs by minimizing the reconstruction loss of the joint state, varying the parameter initialization, learning rate, and number of optimization steps across candidates.
From these candidates, we select the one that maximizes $\bar{l}_{\mathrm{NP}}-\bar{l}_{\mathrm{SM}}$, where $\bar{l}_{\mathrm{NP}}$ and $\bar{l}_{\mathrm{SM}}$ are the mean losses evaluated on the model-selection sample.
The parameters of the selected candidate are then fixed for validation and hardware evaluation.
The index register in Eq.~\eqref{eq:joint} labels the training events, while the six-qubit hardware circuit scores one event at a time.

\section{\label{sec:4}Hardware discrimination}

\begin{figure}[htbp]
\centering
\includegraphics[width=0.68\linewidth]{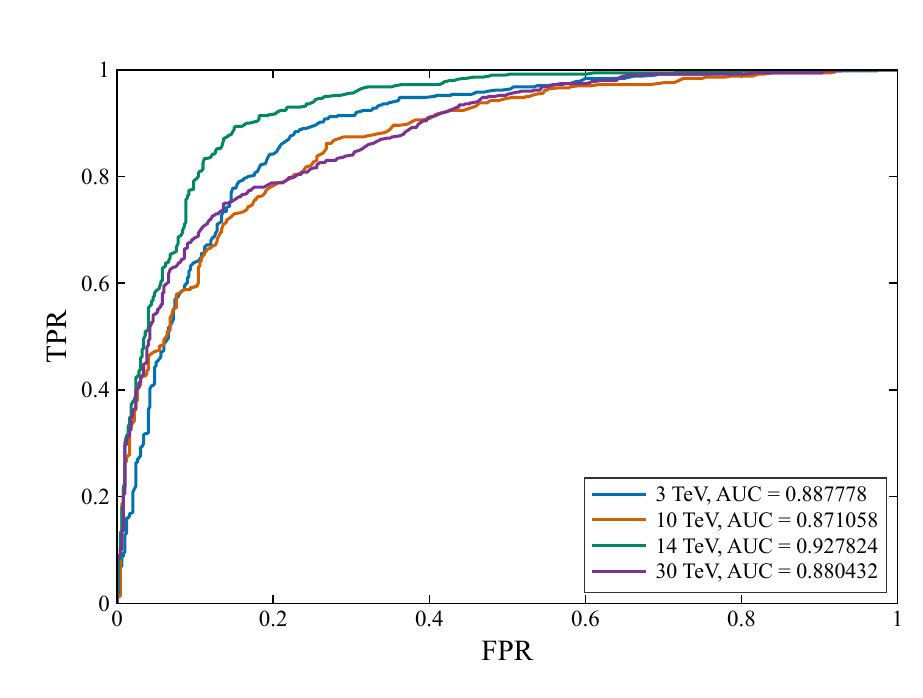}
\caption{Measured ROC curves for the $O_{T_0}$ benchmark on Tianyan176. Each energy uses one random test cohort of $500$ SM and $500$ NP events, with $2000$ shots per event. The false-positive rate is the fraction of SM events above the score threshold, and the true-positive rate is the corresponding NP fraction. Legend values are the point estimates supplied with the original curves.}
\label{fig:roc}
\end{figure}

\begin{figure}[htbp]
\centering
\includegraphics[width=0.64\linewidth]{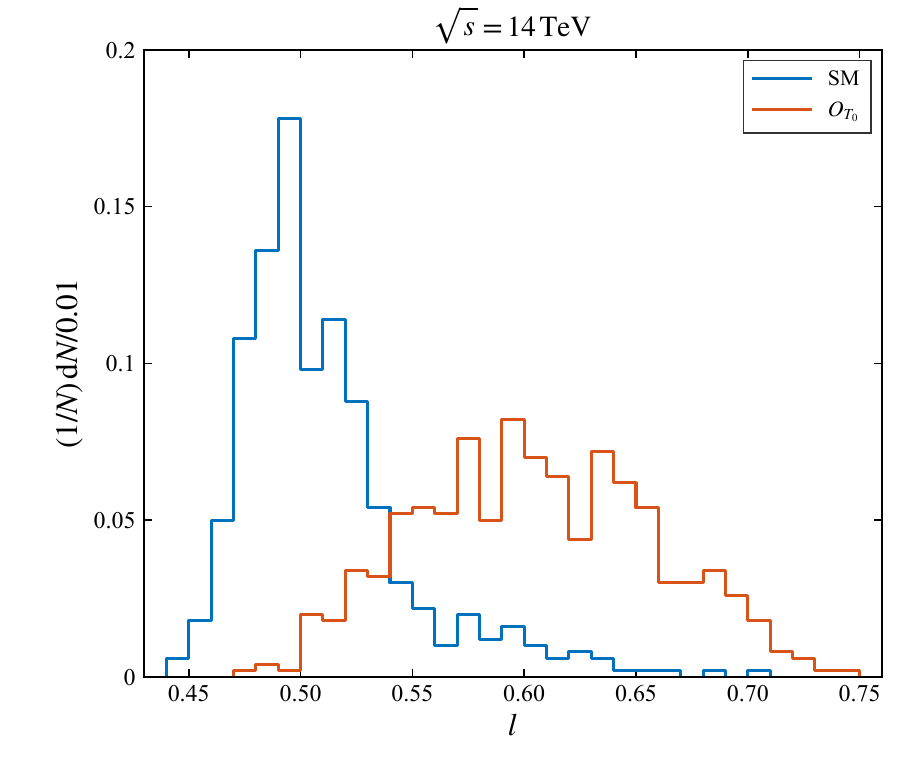}
\caption{Normalized hardware loss distributions at $\sqrt{s}=14~\mathrm{TeV}$ for the same SM and $O_{T_0}$ cohorts used in Fig.~\ref{fig:roc}. Each histogram contains $500$ events. Fig.~\ref{fig:hardwareloss} includes all four energies.}
\label{fig:loss}
\end{figure}

Fig.~\ref{fig:roc} shows the ROC curves obtained by varying the threshold $l_{\rm th}$ in the event selection $\widehat l_n>l_{\rm th}$. The corresponding AUCs are $0.888$, $0.871$, $0.928$, and $0.880$ at $3$, $10$, $14$, and $30~\mathrm{TeV}$, respectively. The AUC equals the probability that a randomly drawn NP event has a higher reconstruction loss than a randomly drawn SM event, plus half the probability that their losses are equal. Fig.~\ref{fig:loss} shows the reconstruction-loss distributions for the $14~\mathrm{TeV}$ hardware sample, which yields the largest measured AUC. The NP distribution is shifted toward higher losses relative to the SM distribution, consistent with the separation seen in the ROC curve. Appendix~\ref{sec:hardware} shows the corresponding distributions at the other three energies and the distributions for the independent validation samples.

\section{\label{sec:5}Connection to collider sensitivity}
With the reconstruction loss threshold fixed, the cross-section after selection for a single nonzero Wilson coefficient $c_i$ is
\begin{equation}
\sigma=\sigma_{\rm SM}+\sigma_{\rm int}c_i+\sigma_{\rm NP}c_i^2,
\label{eq:rate}
\end{equation}
where $\sigma_{\rm SM}$ is the SM contribution, $\sigma_{\rm int}c_i$ denotes the interference between the SM and aQGC, and $\sigma_{\rm NP}c_i^2$ is the aQGC induced contribution.
We compare the expected constraints on the Wilson coefficients obtained using the QAE selection with those reported for conventional kinematic cuts in the same process~\cite{Yang:2020rjt}. The conventional selection requires events to satisfy both $p_{T,\min}/\sqrt{s}>0.12$ and $|\Delta_4|/s^4>3\times10^{-5}$, where $p_{T,\min}$ is the smallest transverse momentum of the three leading photons and $\Delta_4$ is a Gram determinant constructed from the beam and photon four-momenta.

\begin{table}[!htbp]
\caption{Projected coefficient sensitivities at $\sqrt{s}=14$~TeV, integrated luminosity $10~\mathrm{ab}^{-1}$, and statistical signal sensitivity $\mathcal S_{\rm stat}=2$. Endpoints are in $10^{-4}~\mathrm{TeV}^{-4}$. The conventional results are from Table~5 of Ref.~\cite{Yang:2020rjt}.}
\label{table:main-comparison}
\centering
\begin{tabular}{lcc}
\hline\hline
Coefficient & Conventional cuts & QAE selection\\
\hline
$c_0$ & $[-10.14,\,4.09]$  & $[-10.29,\,4.44]$\\
$c_2$ & $[-24.37,\,6.62]$  & $[-24.25,\,7.35]$\\
$c_5$ & $[-2.45,\,0.758]$  & $[-2.48,\,0.834]$\\
$c_7$ & $[-6.17,\,1.17]$   & $[-6.28,\,1.32]$\\
$c_8$ & $[-0.399,\,0.121]$ & $[-0.405,\,0.134]$\\
$c_9$ & $[-1.00,\,0.188]$  & $[-1.05,\,0.214]$\\
\hline\hline
\end{tabular}
\end{table}

The resulting projected intervals are summarized in Table~\ref{table:main-comparison}.
Across the six coefficients, the QAE intervals are approximately $2.0\%$--$6.4\%$ wider than the published cut-based intervals.
Appendices~\ref{sec:scans} and~\ref{sec:projections} give the full coefficient scans and comparisons at the other three collision energies.

A six-qubit QAE on Tianyan176 distinguishes simulated SM and NP events in $\mu^+\mu^-\to\gamma\gamma\gamma$, with measured AUCs ranging from $0.871$ to $0.928$ across four c.m. energies.

\section{\label{sec:6}Conclusions}

The reconstruction loss measured on Tianyan176 distinguishes the NP signals from the SM background in simulated $\mu^+\mu^-\to\gamma\gamma\gamma$ events.
Using only the four-momenta of the three highest-energy reconstructed photons, the six-qubit QAE yields AUCs ranging from $0.871$ to $0.928$ across the four collision energies studied ($3$, $10$, $14$, and $30~\mathrm{TeV}$).

A threshold on this score defines the event selection for the simulated cross-section analysis.
The calculation varies one dimension-8 Wilson coefficient at a time and includes interference between the SM and aQGC amplitudes.
At $\sqrt{s}=14~\mathrm{TeV}$ and an integrated luminosity of $10~\mathrm{ab}^{-1}$, the resulting $\mathcal S_{\rm stat}=2$ intervals for the six coefficients are $2.0\%$--$6.4\%$ wider than those reported for conventional kinematic cuts~\cite{Yang:2020rjt}.
The measured separation between SM and NP signals suggests that quantum hardware could aid in identifying anomalous events in future NP searches at colliders.
\begin{acknowledgments}
We thank the Tianyan quantum computing cloud platform of China Telecom Quantum Information Technology Group Co., LTD., for providing quantum computer time and technical support. This work was supported in part by the National Natural Science Foundation of China under Grants No. 12575106 and 12147214, and by the Basic Scientific Research Project of the Education Department of Liaoning Province under Grant No. LJ212510165024.
\end{acknowledgments}

\appendix
\setcounter{secnumdepth}{1}

\section{Effective-field-theory benchmark}
The dimension-8 transverse operators generating aQGCs are~\cite{Eboli:2006wa,Eboli:2016kko}
\begin{align}
O_{T_0}&={\rm Tr}\!\left[\widehat W_{\mu\nu}\widehat W^{\mu\nu}\right]
          {\rm Tr}\!\left[\widehat W_{\alpha\beta}\widehat W^{\alpha\beta}\right],\\
O_{T_1}&={\rm Tr}\!\left[\widehat W_{\alpha\nu}\widehat W^{\mu\beta}\right]
          {\rm Tr}\!\left[\widehat W_{\mu\beta}\widehat W^{\alpha\nu}\right],\\
O_{T_2}&={\rm Tr}\!\left[\widehat W_{\alpha\mu}\widehat W^{\mu\beta}\right]
          {\rm Tr}\!\left[\widehat W_{\beta\nu}\widehat W^{\nu\alpha}\right],\\
O_{T_5}&={\rm Tr}\!\left[\widehat W_{\mu\nu}\widehat W^{\mu\nu}\right]
          B_{\alpha\beta}B^{\alpha\beta},\\
O_{T_6}&={\rm Tr}\!\left[\widehat W_{\alpha\nu}\widehat W^{\mu\beta}\right]
          B_{\mu\beta}B^{\alpha\nu},\\
O_{T_7}&={\rm Tr}\!\left[\widehat W_{\alpha\mu}\widehat W^{\mu\beta}\right]
          B_{\beta\nu}B^{\nu\alpha},\\
O_{T_8}&=B_{\mu\nu}B^{\mu\nu}B_{\alpha\beta}B^{\alpha\beta},\\
O_{T_9}&=B_{\alpha\mu}B^{\mu\beta}B_{\beta\nu}B^{\nu\alpha}.
\label{eq:operators}
\end{align}
Here $\widehat W\equiv\vec{\sigma}\cdot\vec{W}/2$, with $\sigma$ the Pauli matrices, $W_{\mu\nu}$ and $B_{\mu\nu}$ are the electroweak field-strength tensors. The Lagrangian convention is $\mathcal L=\mathcal L_{\rm SM}+\sum_i c_iO_{T_i}$, with $c_i=f_{T_i}/\Lambda^4$ in $\mathrm{TeV}^{-4}$. The scan switches on one coefficient at a time.
\begin{table}[hbtp]
\centering
\caption{The expected constraints of the $O_{T_{i}}$ coefficients at the LHC, obtained at $95~\%$ C.L.}
\begin{tabular}{c|c|c|c}
\hline
coefficient & constraint & coefficient & constraint \\
\hline
$f_{T_{0}}/\Lambda^4$ &$[-0.12, 0.11]$~\cite{CMS:2019qfk}&$f_{T_{6}}/\Lambda^4$&$[-0.4, 0.4]$~\cite{CMS:2020ypo} \\
$f_{T_{1}}/\Lambda^4$ &$[-0.12, 0.13]$~\cite{CMS:2019qfk}&$f_{T_{7}}/\Lambda^4$&$[-0.9, 0.9]$~\cite{CMS:2020ypo} \\
$f_{T_{2}}/\Lambda^4$ &$[-0.28, 0.28]$~\cite{CMS:2019qfk}&$f_{T_{8}}/\Lambda^4$&$[-0.43, 0.43]$~\cite{CMS:2020fqz} \\
$f_{T_{5}}/\Lambda^4$ &$[-0.5, 0.5]$~\cite{CMS:2020ypo}&$f_{T_{9}}/\Lambda^4$&$[-0.92, 0.92]$~\cite{CMS:2020fqz}\\
\hline
\end{tabular}

\label{table:1}
\end{table}
The expected constraints obtained at the LHC are presented in Table~\ref{table:1}.
\label{sec:operators}
\begin{figure}[htbp]
\begin{center}
\includegraphics[width=0.68\hsize]{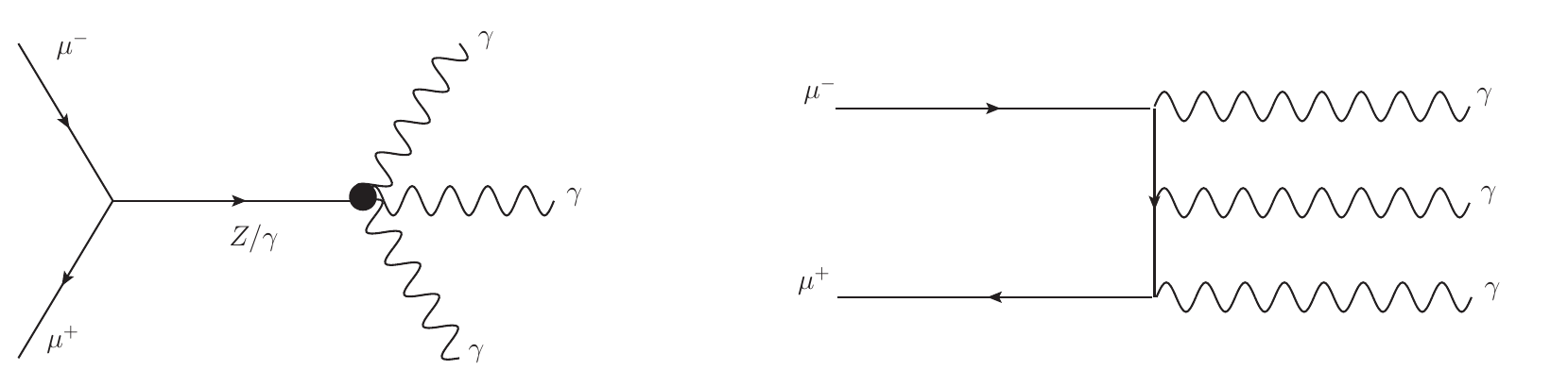}
\caption{\label{fig:Feyman} Feynman diagrams contributing to the tri-photon process $\mu^{+}\mu^{-}\rightarrow\gamma\gamma\gamma$. The left panel displays the diagrams arising from the operators $O_{T_i}$, while the right panel shows a representative SM contribution. The remaining five SM diagrams are obtained by permuting the three photons in the final state.}
\end{center}
\end{figure}
For $\mu^+\mu^-\to\gamma\gamma\gamma$, the Feynman diagrams induced by the $O_{T_i}$ operators are displayed in the left panel of Fig.~\ref{fig:Feyman}, while a representative SM diagram is shown in the right panel.
The remaining five SM diagrams can be obtained by permuting the three final-state photons. The contributions from $O_{T_1}$ and $O_{T_6}$ coincide with those from $O_{T_0}$ and $O_{T_5}$, respectively~\cite{Yang:2020rjt}. The independent scan therefore uses $i=0,2,5,7,8,9$. The hardware measurements use the $O_{T_0}$ benchmark with SMEFT interference excluded. The cross-section scans in Appendix~\ref{sec:scans} include interference.

\section{Event samples and model selection}
\label{sec:samples}
Events are generated using \texttt{MadGraph5\_aMC@NLO}~\cite{Alwall:2014hca} and passed through a muon collider-like \texttt{Delphes} detector simulation~\cite{deFavereau:2013fsa}. The generation requirements are
\begin{equation}
p_{T,\gamma}>10~\mathrm{GeV},\qquad |\eta_\gamma|<2.5,
\qquad\Delta R_{\gamma\gamma}>0.4,
\end{equation}
where $\Delta R_{\gamma\gamma}=\sqrt{(\Delta\eta)^2+(\Delta\phi)^2}$. Events must contain at least three reconstructed photons. The three highest-energy photons are ordered by decreasing energy and their four-momentum components are concatenated into twelve input features. The data preparation is performed by \texttt{MLAnalysis}~\cite{MLAnalysis}.

Four mutually disjoint event samples are used at each c.m. energy of $3$, $10$, $14$, and $30\;\mathrm{TeV}$.
\begin{enumerate}
\item The training sample contains $1024$ SM events used to update the QAE parameters.
\item The model-selection sample contains $10\,000$ SM and $10\,000$ NP events. We select the candidate that maximizes the difference $\bar{l}_{\mathrm{NP}}-\bar{l}_{\mathrm{SM}}$ between the mean losses of the two classes.
\item The independent validation sample contains $10\,000$ events per class and is used to evaluate the selected model.
\item The test pool contains $10\,000$ events per class. A single random sample of 500 SM and 500 NP events is drawn from this pool for hardware evaluation.
\end{enumerate}
Only SM events enter the parameter-update objective. By contrast, model selection evaluates each candidate parameter vector $\boldsymbol\theta$ using
\begin{equation}
\Delta_{\rm sel}(\boldsymbol\theta)=
\frac{1}{N_{\rm NP,sel}}\sum_{n\in\mathrm{NP,sel}}l_n(\boldsymbol\theta)
-\frac{1}{N_{\rm SM,sel}}\sum_{n\in\mathrm{SM,sel}}l_n(\boldsymbol\theta).
\label{eq:selection}
\end{equation}
The candidate that maximizes $\Delta_{\rm sel}$ is retained, and its parameters are then held fixed for validation and hardware evaluation.

\section{Quantum reconstruction and its measurement}
\label{sec:quantum}
\begin{figure}[htbp]
\centering
\includegraphics[width=0.94\textwidth]{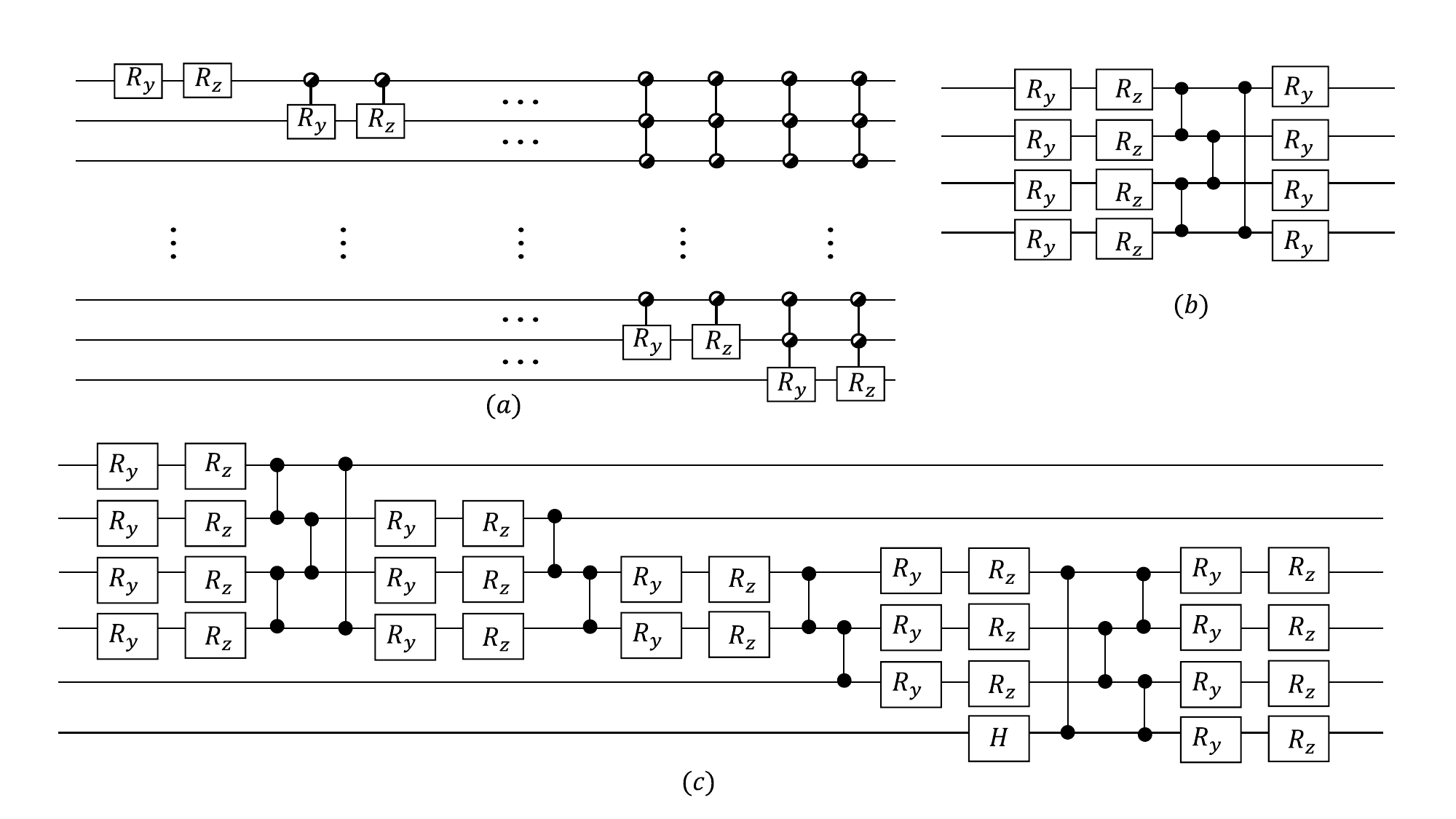}
\caption{Quantum circuits for training and event reconstruction. Amplitude encoding of the joint training state using uniformly controlled rotations is depicted in (a), the four-qubit data-encoding circuit $U_H(\mathbf{x}_n)$ for an individual event is shown in (b), and (c) shows the six-qubit QAE circuit $U_A(\boldsymbol\theta)$ for state compression and reconstruction. The schematic gate pattern is distinct from the compiled native-gate circuit executed on the processor.}
\label{fig:circuits}
\end{figure}

The four-qubit event state is $|\psi_n\rangle=U_H(\mathbf x_n)|0000\rangle$, where $U_H$ is the hardware-efficient encoding in Fig.~\ref{fig:circuits}(b). Fig.~\ref{fig:circuits}(c) shows the QAE circuit. It includes four input qubits and two additional qubits initialized in $\lvert00\rangle$.

Let $R$ denote the four reconstructed output qubits and $D$ the two discarded qubits. With $U_A(\boldsymbol\theta)$ the ideal six-qubit QAE operation, define
\begin{equation}
\rho^{\rm rec}_n=\operatorname{Tr}_{D}\!\left[
U_A(\boldsymbol\theta)\bigl(|\psi_n\rangle\langle\psi_n|\otimes|00\rangle\langle00|\bigr)
U_A^\dagger(\boldsymbol\theta)\right].
\label{eq:channel}
\end{equation}
The output generally need not be pure. The reconstruction fidelity relative to the pure encoded input and the score are
\begin{equation}
f_n=\langle\psi_n|\rho^{\rm rec}_n|\psi_n\rangle,
\qquad l_n=1-f_n.
\label{eq:fidelity}
\end{equation}
The pure-state formula $f_n=|\langle\psi_n|\phi_n\rangle|^2$ is the special case $\rho^{\rm rec}_n=|\phi_n\rangle\langle\phi_n|$. Applying the inverse encoder to register $R$ gives
\begin{equation}
\Pr(0000\text{ on }R)=
\langle0000|U_H^\dagger\rho^{\rm rec}_nU_H|0000\rangle=f_n.
\label{eq:measurement}
\end{equation}
Note that this probability is marginalized over $D$ and does not correspond to the probability of the full six-qubit state being $|000000\rangle$. Fig.~\ref{fig:circuit} shows the overlap measurement on the reconstructed register.

For the optimization construction, the $M=1024$ training states are represented by
\begin{equation}
|\Psi\rangle=\frac{1}{\sqrt M}\sum_{n=0}^{M-1}|\psi_n\rangle\otimes|n\rangle.
\label{eq:joint-appendix}
\end{equation}
The index $n$ in Eq.~\eqref{eq:joint-appendix} requires ten qubits in the joint training representation for $M=1024$ states. During hardware evaluation, the six-qubit circuit scores one event at a time.

\section{Validation and hardware results}
\label{sec:hardware}
\begin{figure}[htbp]
\centering
\includegraphics[width=0.52\textwidth]{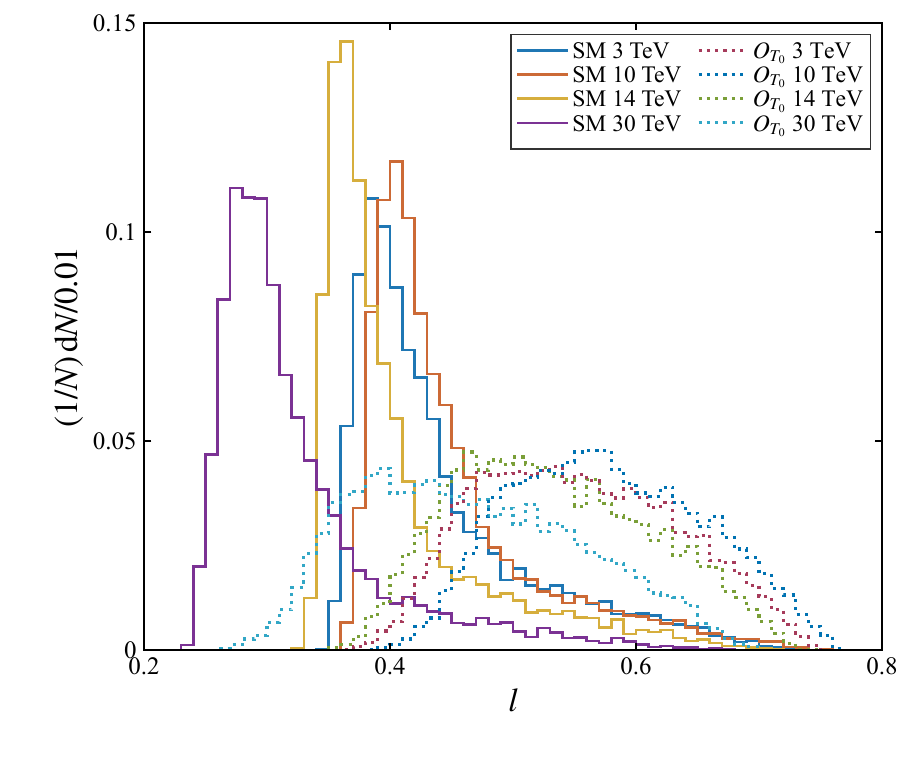}
\caption{The normalized distributions of $l$ for the SM and the NP signals at $3$, $10$, $14$, and $30\;\mathrm{TeV}$. The independent validation sample contains $10\,000$ events per class at each energy.}
\label{fig:validation}
\end{figure}

Fig.~\ref{fig:validation} shows the separation between the SM background and NP signal distributions in the validation sample.

Each hardware circuit uses six qubits and $N_{\rm shot}=2000$ shots on Tianyan176. The number of $0000$ outcomes on the four reconstructed outputs determines the score through
\begin{equation}
\widehat l_n=1-\frac{n_{0000}}{2000}.
\label{eq:estimator}
\end{equation}
We use the raw counts without error mitigation or postselection.

\begin{figure}[htbp]
\centering
\includegraphics[width=0.48\textwidth]{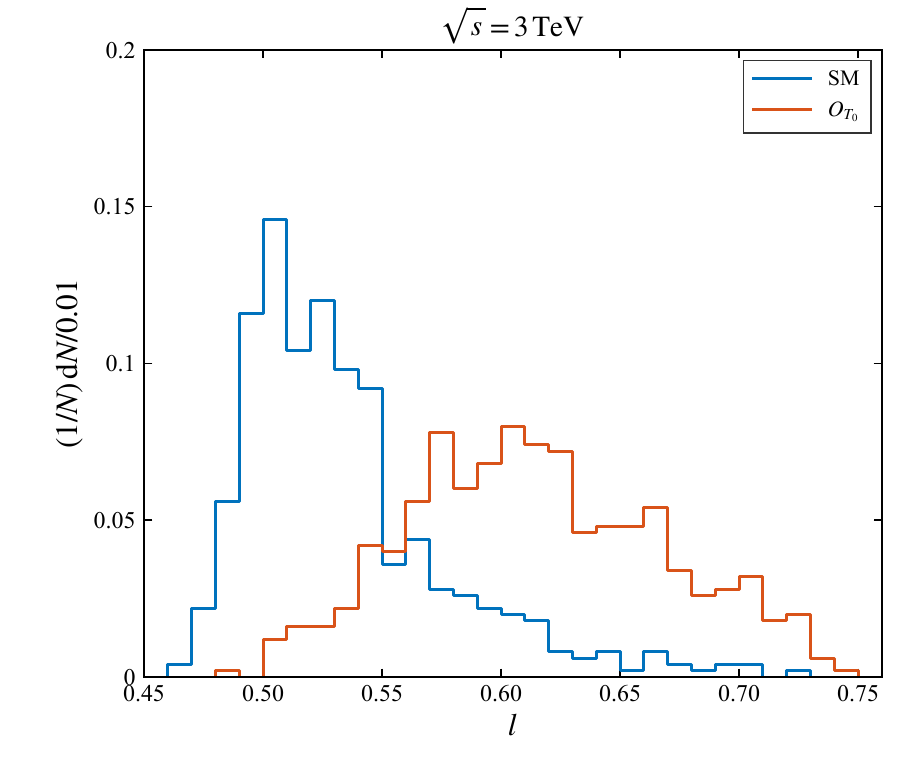}\hfill
\includegraphics[width=0.48\textwidth]{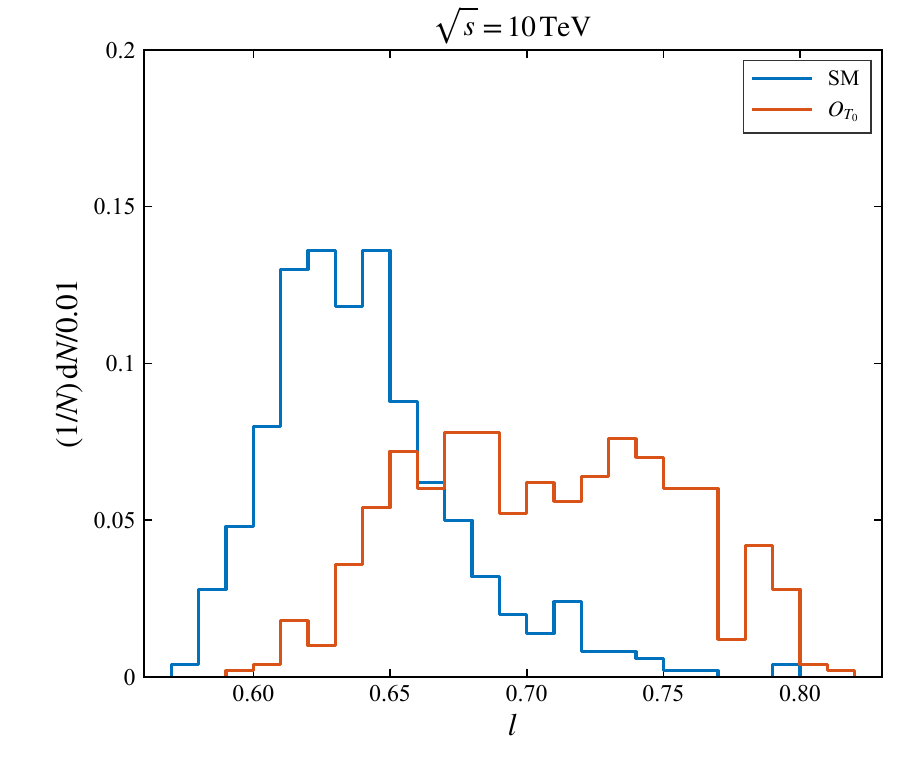}\\[1ex]
\includegraphics[width=0.48\textwidth]{max_auc_loss_14TeV.pdf}\hfill
\includegraphics[width=0.48\textwidth]{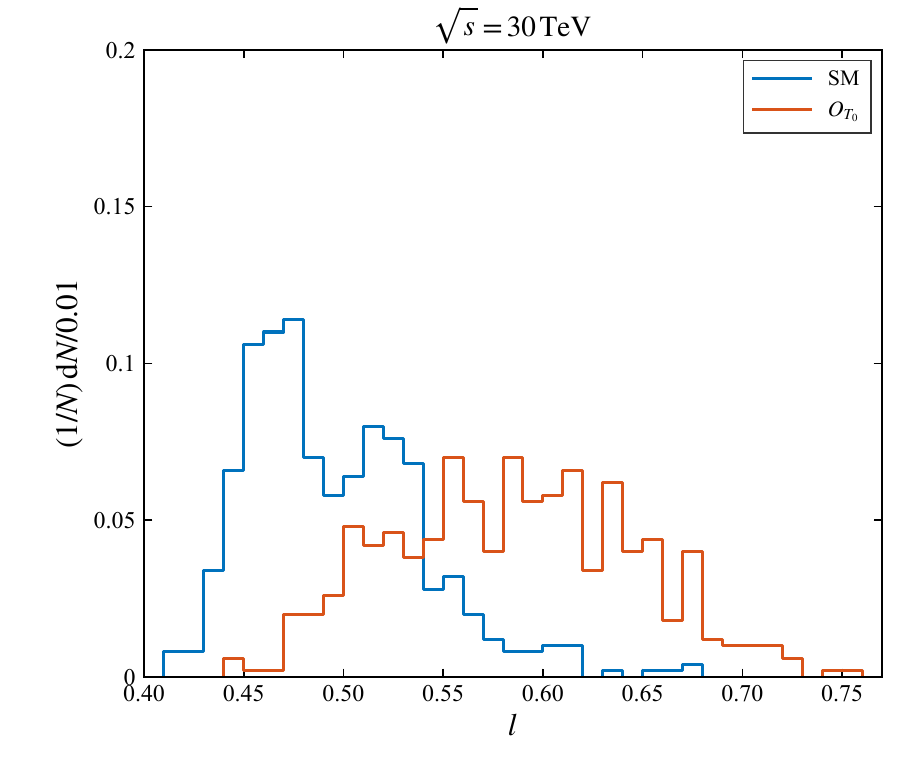}
\caption{Hardware loss distributions for SM and $O_{T_0}$ events at $3$ and $10\;\mathrm{TeV}$ (upper row), and $14\;\mathrm{TeV}$ and $30\;\mathrm{TeV}$ (lower row). Each class contains $500$ events and each circuit uses $2000$ shots. The bin width is $0.01$. Fig.~\ref{fig:roc} uses these same cohorts.}
\label{fig:hardwareloss}
\end{figure}

The measured AUCs at $3$, $10$, $14$, and $30\;\mathrm{TeV}$ are $0.887778$, $0.871058$, $0.927824$, and $0.880432$, respectively. Fig.~\ref{fig:hardwareloss} shows the loss distributions for these four test cohorts.

\clearpage
\section{Selected cross-sections and coefficient scans}
\label{sec:scans}
\begin{figure}[p]
\centering
\includegraphics[width=0.325\linewidth]{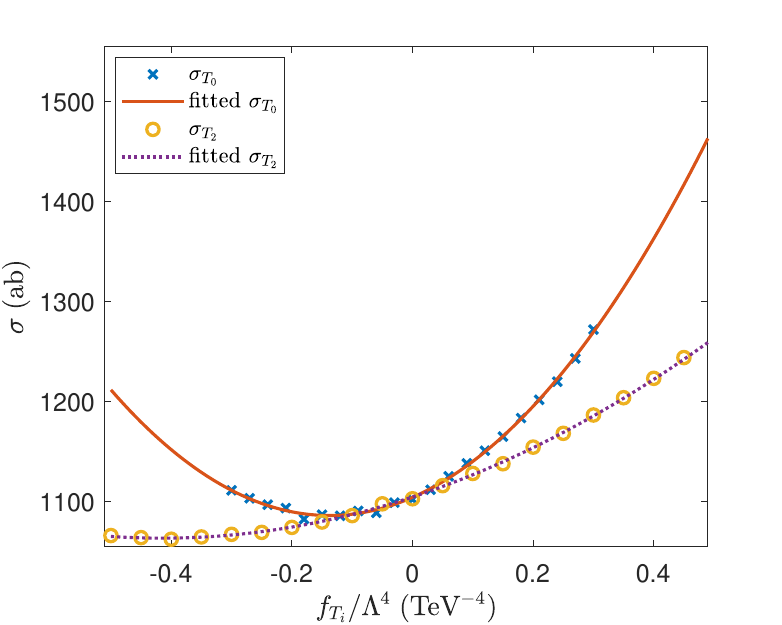}\hfill
\includegraphics[width=0.325\linewidth]{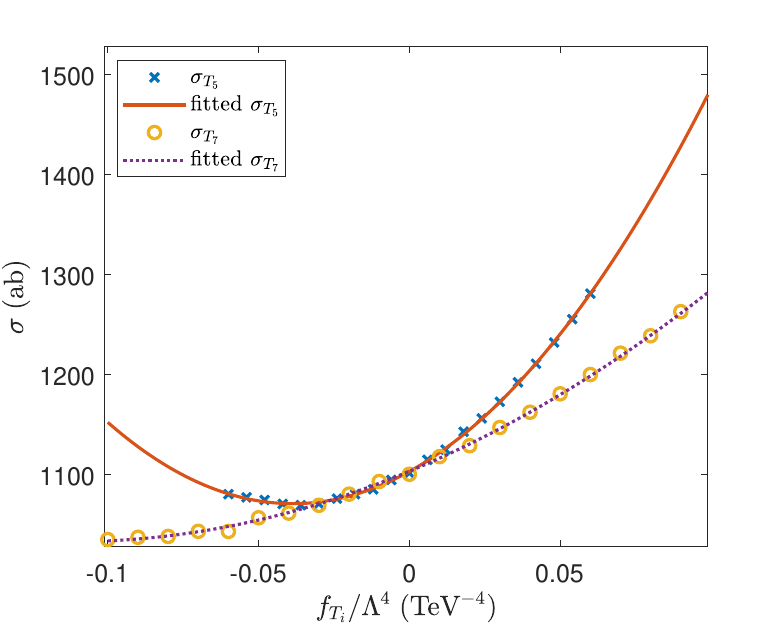}\hfill
\includegraphics[width=0.325\linewidth]{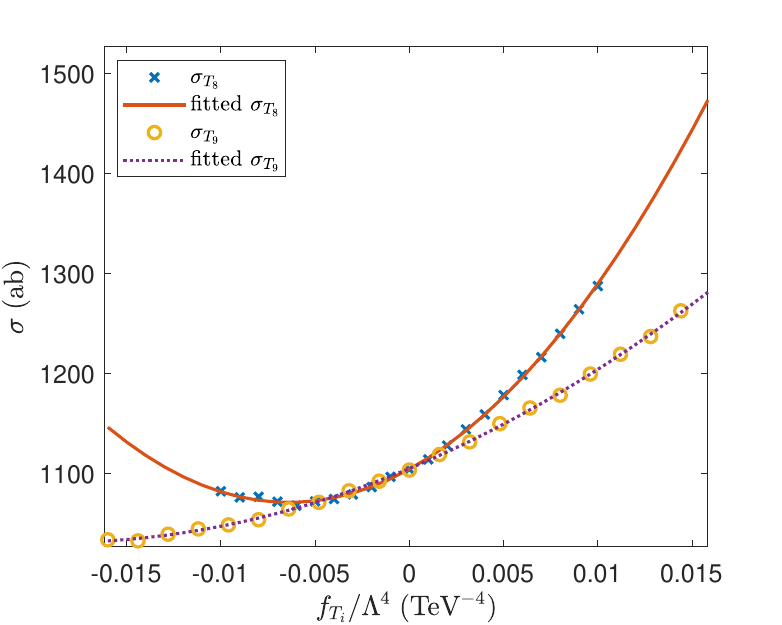}\\[0.6ex]
\includegraphics[width=0.325\linewidth]{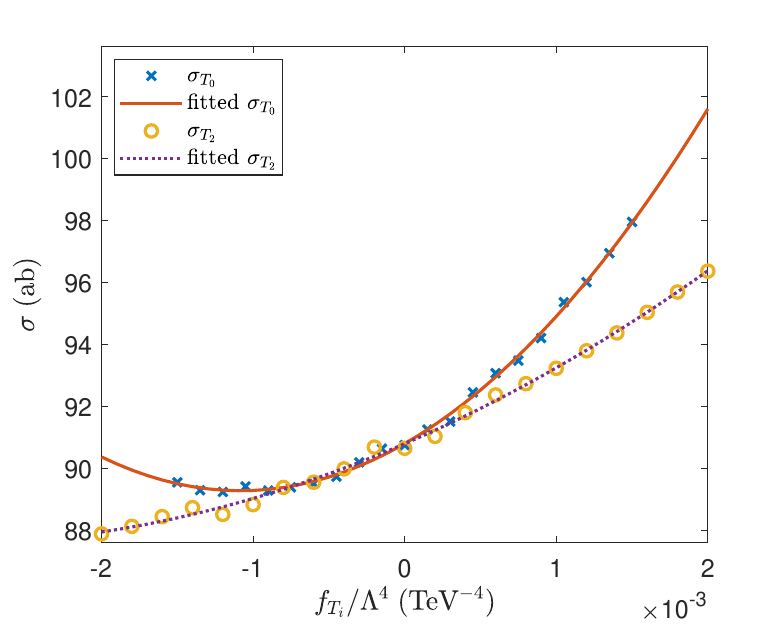}\hfill
\includegraphics[width=0.325\linewidth]{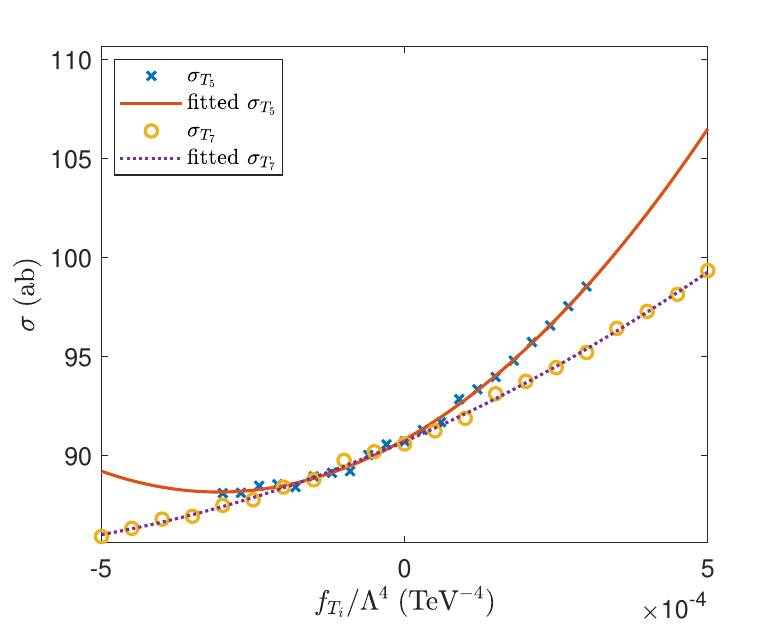}\hfill
\includegraphics[width=0.325\linewidth]{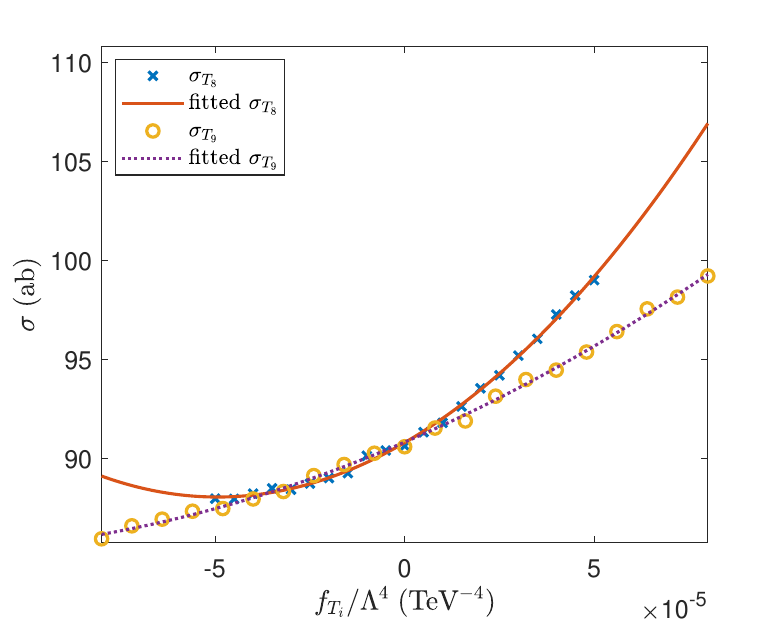}\\[0.6ex]
\includegraphics[width=0.325\linewidth]{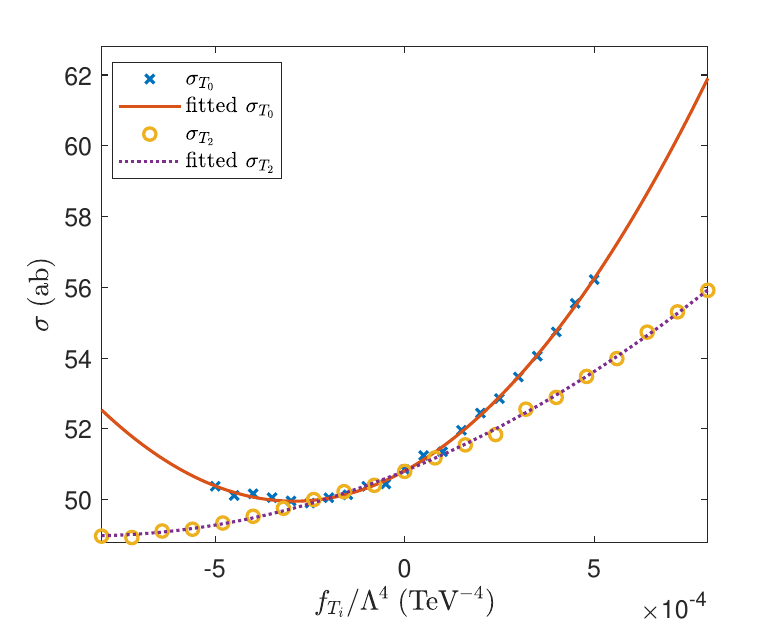}\hfill
\includegraphics[width=0.325\linewidth]{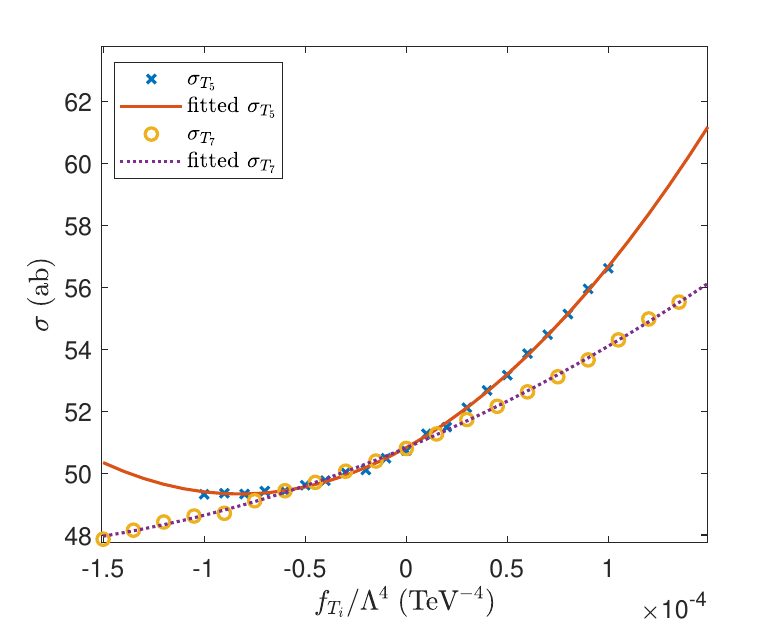}\hfill
\includegraphics[width=0.325\linewidth]{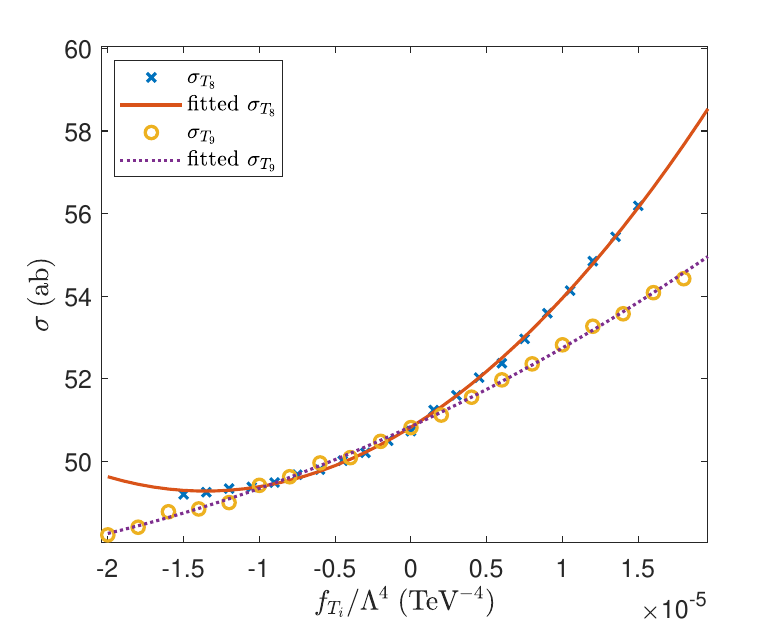}\\[0.6ex]
\includegraphics[width=0.325\linewidth]{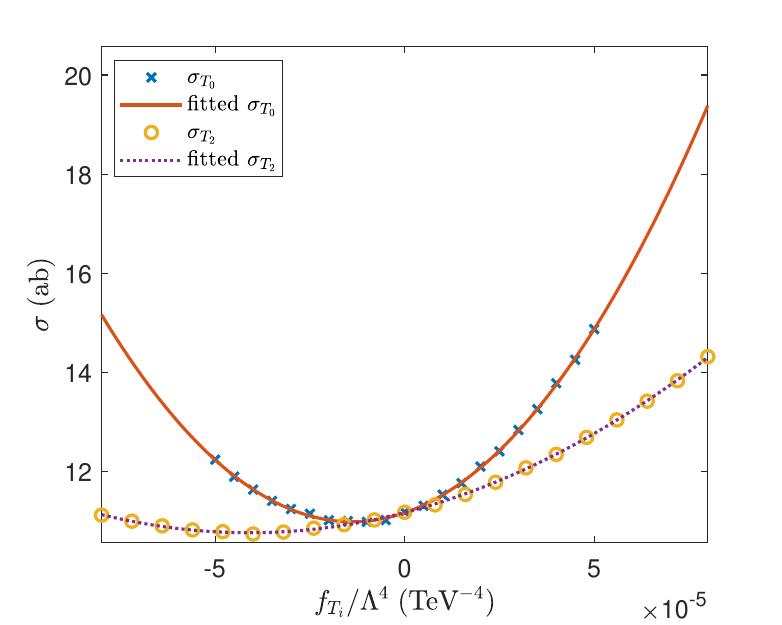}\hfill
\includegraphics[width=0.325\linewidth]{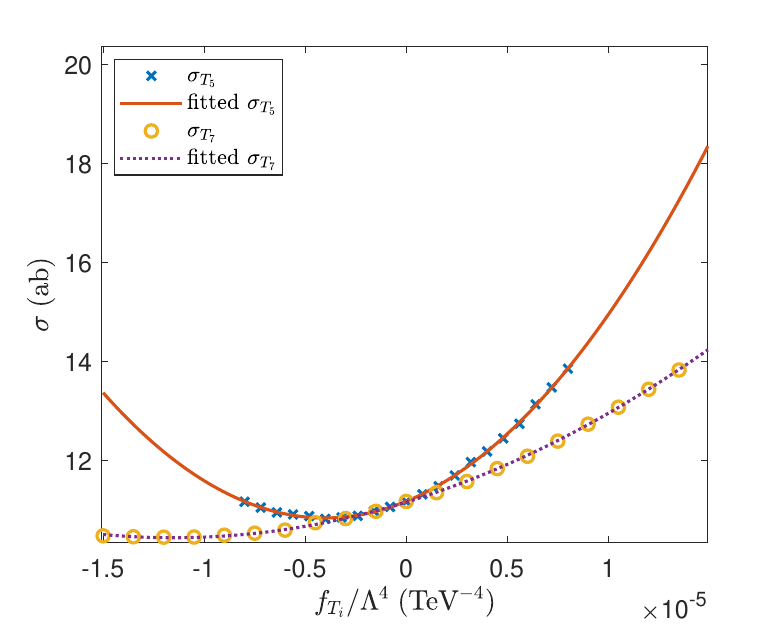}\hfill
\includegraphics[width=0.325\linewidth]{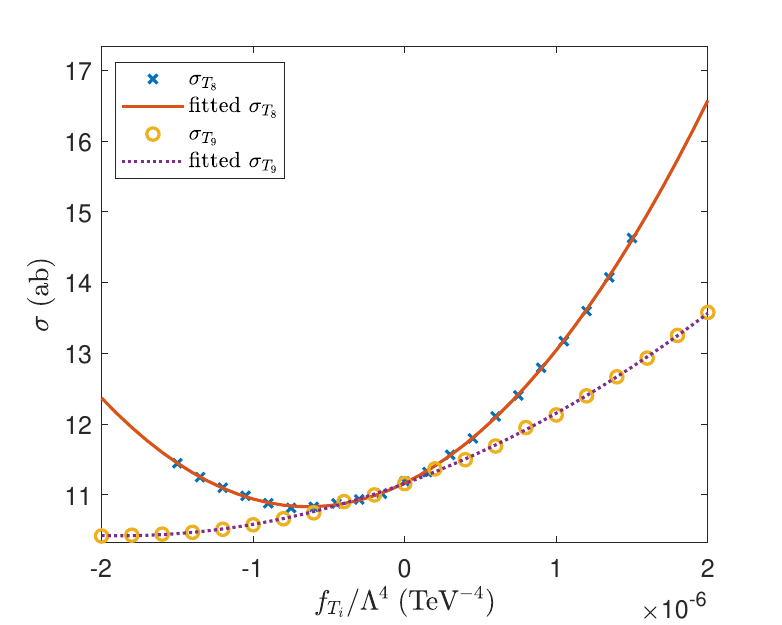}
\caption{Comparison of the cross-sections $\sigma$ after cuts with the fitted $\sigma$ using bilinear functions of $f_{T_i}/\Lambda^4$ at different collision energies. Rows 1 to 4 correspond to $\sqrt{s} = 3, 10, 14$, and $30\;\mathrm{TeV}$, respectively.}
\label{fig:fits}
\end{figure}

With a fixed selection $l>l_{\rm th}$ and one active coefficient $c_i$, the accepted cross-section has the quadratic form
\begin{equation}
\sigma(c_i)=\sigma_{\rm SM}+\sigma_{\rm int}c_i+\sigma_{\rm NP}c_i^2.
\label{eq:fit}
\end{equation}
The linear term accounts for SMEFT interference. Ref~\cite{Yang:2020rjt} gives analytic expressions before event selection. Table~\ref{table:scan} lists the coefficient scan ranges. Fig.~\ref{fig:fits} shows the simulated cross-sections after QAE selection and the corresponding quadratic fits.

\begin{table}[htbp]
\centering
\small
\setlength{\tabcolsep}{4pt}
\begin{tabular}{c|c|c|c|c} 
\hline
 $\sqrt{s}$ & 3\;{\rm TeV} &10\;{\rm TeV} & 14\;{\rm TeV} & 30\;{\rm TeV} \\ 
 \hline
 Unit of coefficient & $ ({\rm TeV}^{-4})$ & $(10^{-3}\;{\rm TeV}^{-4})$ &$ (10^{-3}\;{\rm TeV}^{-4})$&$(10^{-4}\;{\rm TeV}^{-4})$\\
\hline
 $f_{T_0}/\Lambda ^4$ &[-0.3, 0.3]&[-1.5, 1.5] &[-0.5, 0.5]& [-0.5, 0.5]\\ 
\hline
 $f_{T_2}/\Lambda ^4$ & [-0.5, 0.5]&[-2.0, 2.0] & [-0.8, 0.8]& [-0.8, 0.8]\\
\hline
 $f_{T_5}/\Lambda ^4$ & [-0.06, 0.06] &[-0.3, 0.3]& [-0.1, 0.1]& [-0.08, 0.08]\\
\hline
 $f_{T_7}/\Lambda ^4$ & [-0.1, 0.1]& [-0.5, 0.5] &[-0.15, 0.15] &[-0.15, 0.15] \\
\hline
 $f_{T_8}/\Lambda ^4$ &[-0.01, 0.01]& [-0.05, 0.05]& [-0.015, 0.015]& [-0.015, 0.015] \\ [1ex]
\hline
 $f_{T_9}/\Lambda ^4$ &[-0.016, 0.016]& [-0.08, 0.08]& [-0.02, 0.02]& [-0.02, 0.02] \\ [1ex] 
\hline
\end{tabular}
\caption{Coefficient scan ranges.}
\label{table:scan}
\end{table}

\clearpage
\section{Statistical signal significance and expected coefficient constraints}
\label{sec:projections}
The statistical signal significance is calculated, which is defined as~\cite{Cowan:2010js}
\begin{equation}
\begin{gathered}
\mathcal S_{\rm stat}=\sqrt{2\left[(N_{\rm bg}+N_s)
\ln\left(1+\frac{N_s}{N_{\rm bg}}\right)-N_s\right]},\\
N_s=(\sigma-\sigma_{\rm SM})L,\qquad N_{\rm bg}=\sigma_{\rm SM}L.
\end{gathered}
\label{eq:significance}
\end{equation}
We consider the luminosities for both the ``conservative'' and ``optimistic'' scenarios~\cite{AlAli:2021let}. The resulting constraints on the coefficients at $\mathcal S_{\rm stat}=2,\;3,\;5$ are listed in Tables~\ref{table:conservative} and~\ref{table:optimistic}, respectively.

\begin{table}[!htbp]
\centering
\small
\setlength{\tabcolsep}{4pt}
\renewcommand{\arraystretch}{0.76}
\begin{tabular}{c|c|c|c|c|c} 
\hline
  &  &3\;{\rm TeV} &10\;{\rm TeV} &14\;{\rm TeV} &30\;{\rm TeV} \\
  &$\mathcal S_{\rm stat}$& $1\;{\rm ab}^{-1}$&$10\;{\rm ab}^{-1}$&$10\;{\rm ab}^{-1}$ &$10\;{\rm ab}^{-1}$ \\
  &  &$(10^{-2}\;{\rm TeV^{-4}})$&$(10^{-4}\;{\rm TeV^{-4}})$&$(10^{-4}\;{\rm TeV^{-4}})$&$(10^{-5}\;{\rm TeV^{-4}})$ \\
\hline
  & 2 &[-43.62, 16.07]&[-35.06,13.40]& [-10.29, 4.44]&[-6.26, 3.59]\\
$f_{T_{0}}(f_{T_{1}})/\Lambda^4$& 3 &[-49.08, 21.53]&[-39.56, 17.90]& [-11.73, 5.88]&[-7.44, 4.67] \\
  & 5 & [-58.13, 30.58]&[-47.03, 25.37]&[-14.11, 8.26]&[-9.22, 6.46]\\
\hline
 & 2 &[-108.87, 25.81]&[-83.19,21.44]&[-24.25, 7.35]&[-14.46, 6.24] \\
$f_{T_{2}}/\Lambda^4$& 3 &[-118.73, 35.68]&[-91.23, 29.48]&[-26.91, 10.00]&[-16.52, 8.31] \\
  & 5 &[-135.62, 52.57]&[-104.93, 43.18]&[-31.40, 14.49]&[-20.00, 11.78] \\
\hline
 & 2 &[-10.64, 2.92]&[-8.64, 2.50]&[-2.48, 0.834]& [-1.50, 0.696]\\
$f_{T_{5}}(f_{T_{6}})/\Lambda^4$& 3 &[-11.72, 4.00]&[-9.55, 3.42]&[-2.77, 1.13]&[-1.72, 0.920] \\
  & 5 &[-13.55, 5.82]&[-11.90, 4.96]&[-3.26, 1.62]&[-2.10, 1.30] \\
\hline
& 2 &[-26.96, 4.47]&[-20.99,3.77]&[-6.28,1.32]& [-3.46, 1.16]\\
$f_{T_{7}}/\Lambda^4$& 3 &[-28.80, 6.30]&[-22.52, 5.30]&[-6.80, 1.84]&[-3.88, 1.58] \\
  & 5 &[-32.03, 9.54]&[-25.19, 7.98]&[-7.70, 2.75]& [-4.59, 2.28] \\
\hline
 &2 &[-1.75, 0.471]&[-1.38, 0.392]&[-0.405,  0.134]&[-0.238, 0.111] \\
$f_{T_{8}}/\Lambda^4$& 3 &[-1.92, 0.646]&[-1.52, 0.536]&[-0.453, 0.181]&[-0.274, 0.147] \\
  & 5 &[-2.22, 0.942]&[-1.77, 0.778]&[-0.532,  0.261]& [-0.334, 0.206]\\
\hline
 &2 &[-4.50, 0.719]&[-3.37, 0.608]&[-1.05, 0.214]&[-0.562, 0.186] \\
$f_{T_{9}}/\Lambda^4$& 3 &[-4.80, 1.02]&[-3.61, 0.855]&[-1.13, 0.299]& [-0.629, 0.253]\\
  & 5 &[-5.32, 1.54]&[-4.04, 1.29]&[-1.28, 0.447]&[-0.743, 0.367]\\
\hline
\end{tabular}
\caption{Projected sensitivities to the $O_{T_i}$ operator coefficients at muon colliders, evaluated for different c.m. energies and integrated luminosities in the ``conservative'' scenario.}
\label{table:conservative}
\end{table}

\begin{table}[!htbp]
\centering
\small
\setlength{\tabcolsep}{4pt}
\renewcommand{\arraystretch}{0.76}
\begin{tabular}{c|c|c|c} 
\hline
  &  &14\;{\rm TeV} &30\;{\rm TeV} \\
  &$\mathcal S_{\rm stat}$& $20\;{\rm ab}^{-1}$&$90\;{\rm ab}^{-1}$ \\
  &  &$(10^{-5}\;{\rm TeV^{-4}})$&$(10^{-6}\;{\rm TeV^{-4}})$ \\
\hline
  & 2 &[-93.06, 34.57]&[-44.43, 16.78] \\
$f_{T_{0}}(f_{T_{1}})/\Lambda^4$& 3 &[-104.77, 46.29]&[-50.10, 22.44] \\
  & 5 & [-124.22,65.73]& [-59.48, 31.82] \\
\hline
 & 2 &[-224.87, 55.79]&[-109.17, 27.01] \\
$f_{T_{2}}/\Lambda^4$& 3 &[-245.97, 76.89]&[-119.39, 37.24] \\
  & 5 &[-281.99, 112.92]& [-136.85, 54.70] \\
\hline
 & 2 &[-22.82, 6.38]&[-11.09, 3.07] \\
$f_{T_{5}}(f_{T_{6}})/\Lambda^4$& 3 &[-25.17, 8.72]&[-12.22, 4.20] \\
  & 5 & [-29.13, 12.68]&[-14.13, 6.11] \\
\hline
& 2 &[-18.32, 9.83]&[-13.60, 4.72] \\
$f_{T_{7}}/\Lambda^4$& 3 &[-63.42, 13.88]&[-29.70, 6.66] \\
  & 5 &[-70.55, 21.01]& [-33.10, 10.06] \\
\hline
 &2 &[-3.74, 1.02]&[-1.76, 0.489] \\
$f_{T_{8}}/\Lambda^4$& 3 &[-4.12, 1.40]&[-1.94, 0.669] \\
  & 5 &[-4.76, 2.04]& [-2.25, 0.974] \\
\hline
 &2 &[-2.78, 1.59]&[-2.33, 0.755] \\
$f_{T_{9}}/\Lambda^4$& 3 &[-10.60, 2.25]&[-4.83, 1.07] \\
  & 5 &[-11.77, 3.41]&[-5.37, 1.61] \\
\hline
\end{tabular}
\caption{Same as Table~\ref{table:conservative}, but for the ``optimistic'' case.}
\label{table:optimistic}
\end{table}

\begin{figure}[p]
\centering
\includegraphics[width=0.47\linewidth]{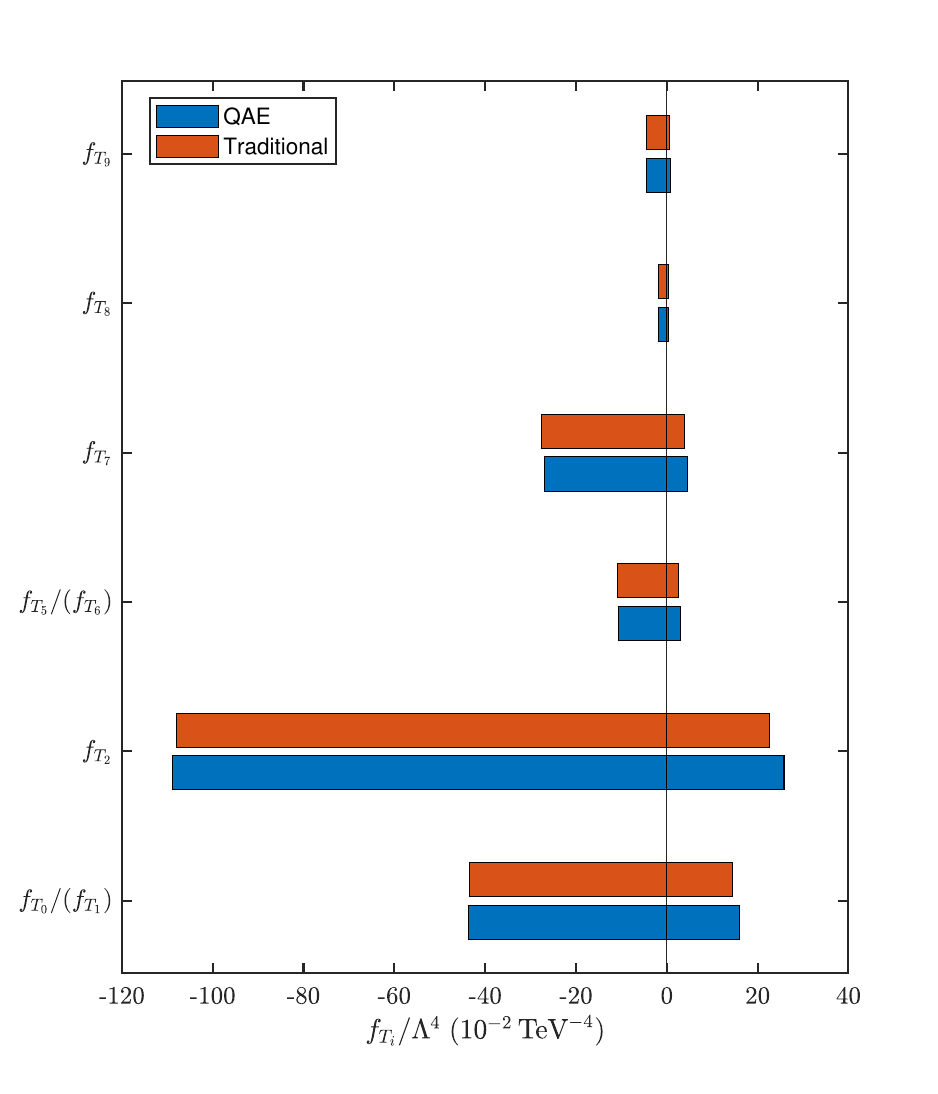}\hfill
\includegraphics[width=0.47\linewidth]{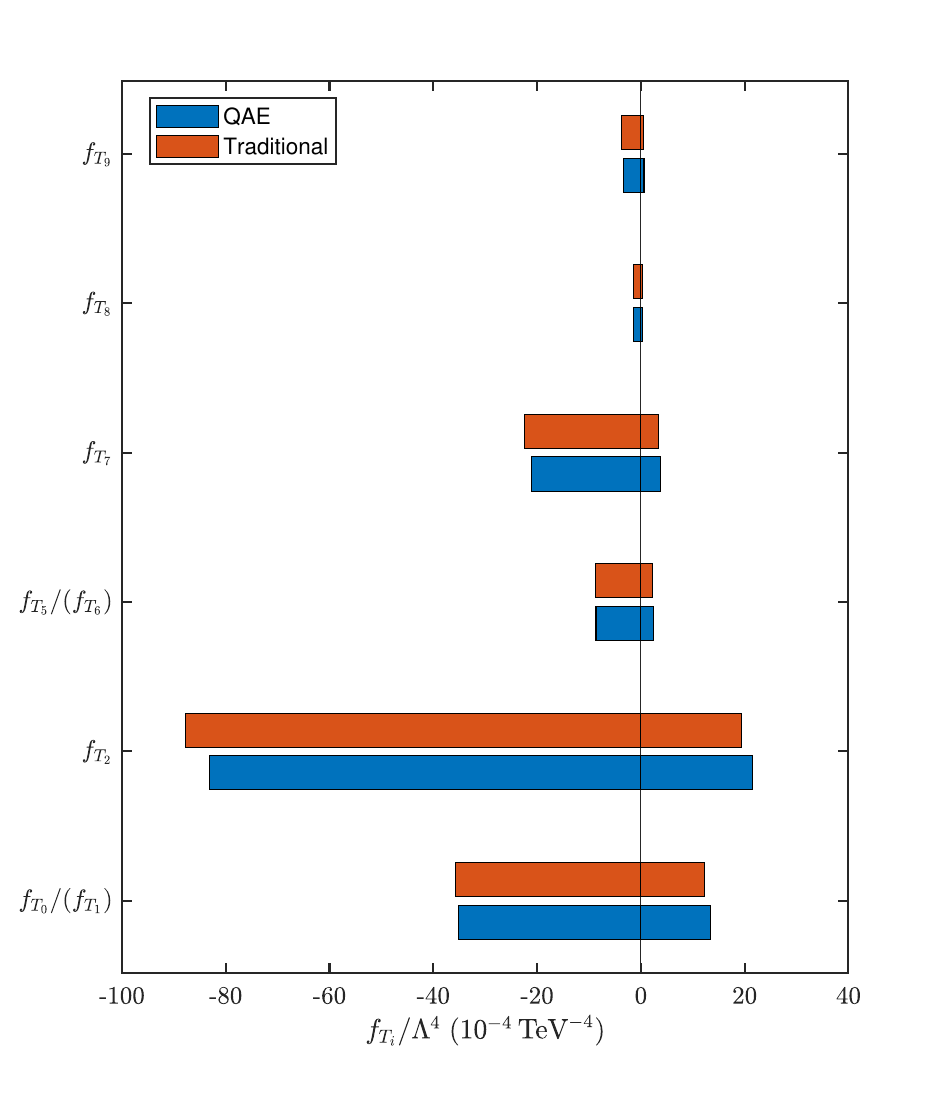}\\[0.6ex]
\includegraphics[width=0.47\linewidth]{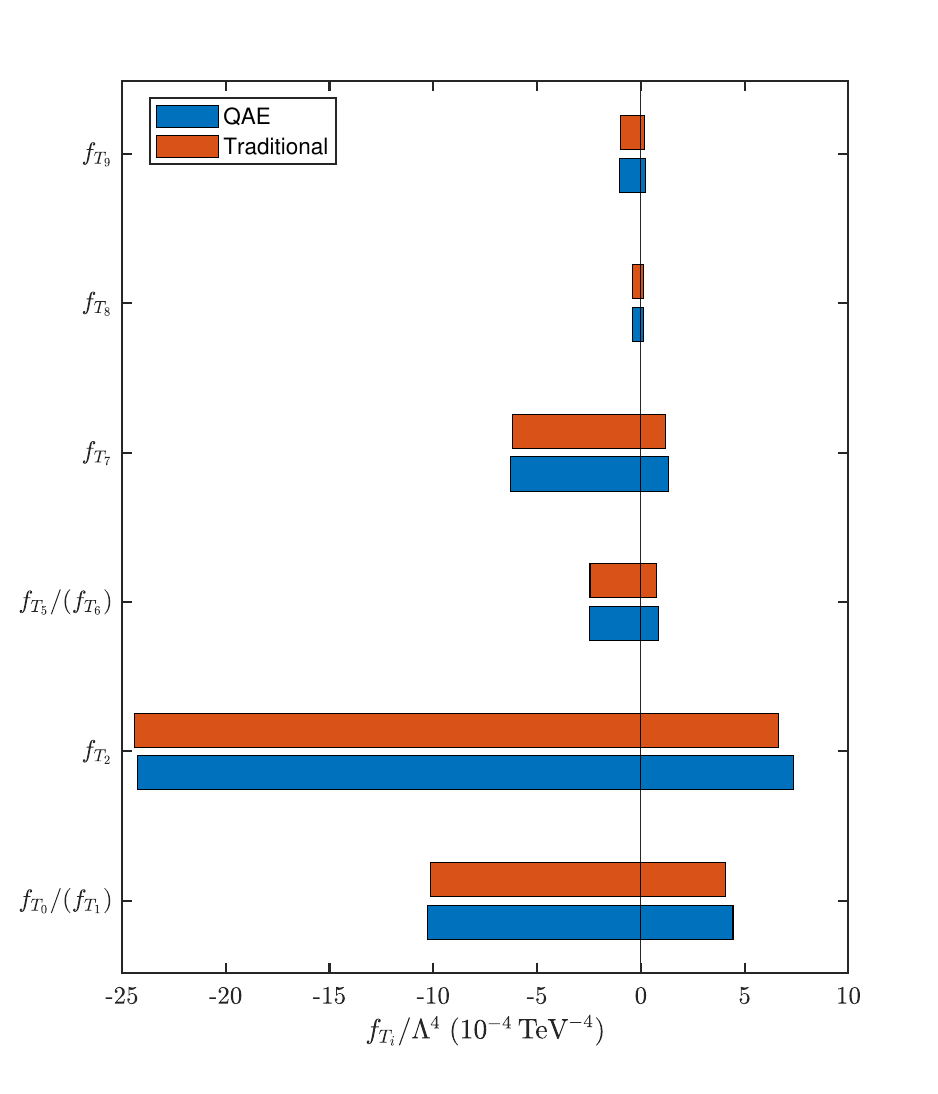}\hfill
\includegraphics[width=0.47\linewidth]{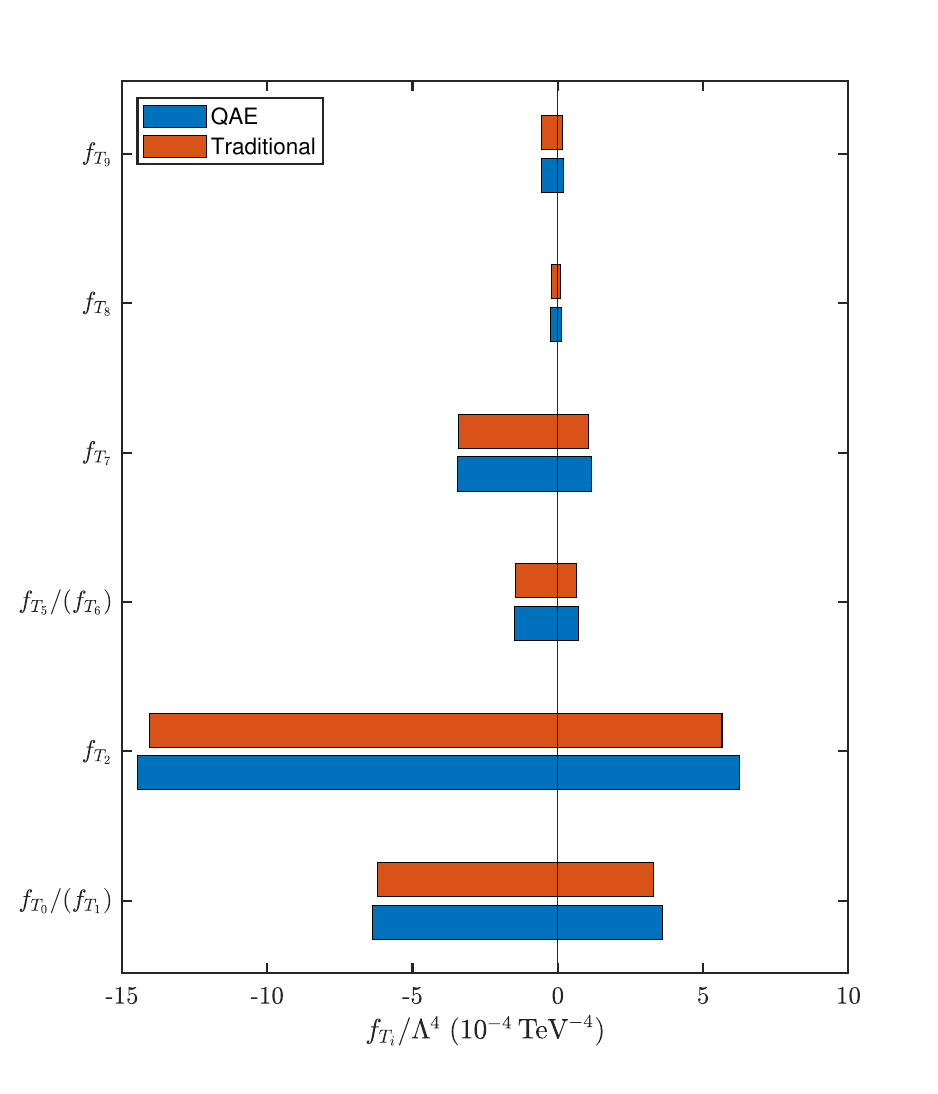}
\caption{Comparison of the expected coefficient constraints obtained for different methods in the ``conservative'' scenario. From left to right, the upper and lower panels correspond to $\sqrt{s}=3,\;10\;\mathrm{TeV}$ and $14,\;30\;\mathrm{TeV}$, respectively.}
\label{fig:compare}
\end{figure}

Fig.~\ref{fig:compare} compares the expected constraints of QAE with the published kinematic analysis~\cite{Yang:2020rjt} using the ``conservative'' luminosity scenarios. The two analyses yield intervals of comparable width.

\clearpage
\bibliography{references}
\end{document}